\documentclass[fleqn,usenatbib]{mnras}

\usepackage[T1]{fontenc}
\usepackage{ae,aecompl}

\usepackage{graphicx}	
\usepackage{amsmath}	
\usepackage{amssymb}	

\usepackage[normalem]{ulem}
\usepackage{color}

\usepackage{newtxtext,newtxmath}

\title[Detecting General Relativistic Orbital Precession in Transiting Hot Jupiters]{Detecting General Relativistic Orbital Precession in Transiting Hot Jupiters}

\author[Antoniciello et al.]{G. Antoniciello$^{1}$ \thanks{E-mail: giuliano.antoniciello@phd.unipd.it}, L. Borsato$^{2}$, G. Lacedelli$^{1,2}$, V. Nascimbeni$^{2}$, O. Barrag\'an$^{3}$, \and R. Claudi$^{2}$
\\
\\
$^{1}$Dipartimento di Fisica e Astronomia "Galileo Galilei", Universit\'a di Padova, Vicolo dell'Osservatorio 3, 35122, Padova, Italy \\
$^{2}$INAF - Osservatorio Astronomico di Padova, Vicolo dell'Osservatorio 5, 35122 Padova, Italy \\
$^{3}$Sub-department of Astrophysics, Department of Physics, University of
Oxford, Oxford, OX1 3RH, UK
}

\date{Accepted XXX. Received YYY; in original form ZZZ}

\pubyear{2021}

\begin{document}
\label{firstpage}
\pagerange{\pageref{firstpage}--\pageref{lastpage}}

\maketitle

\begin{abstract}
Both classical and relativistic weak-field and slow-motion perturbations to planetary orbits can be treated as perturbative corrections to the Keplerian model. In particular, tidal forces and General Relativity (GR) induce small precession rates of the apsidal line. Accurate measurements of these effects in transiting exoplanets could be used to test GR and to gain information about the planetary interiors. Unfortunately, models for transiting planets have a high degree of degeneracy in the orbital parameters that, combined to the uncertainties of photometric transit observations, results in large errors on the determinations of the argument of periastron and precludes a direct evaluation of the apsidal line precession. Moreover, tidal and GR precession time-scales are many order of magnitudes larger than orbital periods, so that on the observational time-spans required to cumulate a precession signal enough strong to be detected, even small systematic errors in transit ephemerides add up to cancel out the tiny variations due to precession. Here we present a more feasible solution to detect tidal and GR precession rates through the observation of variations of the time interval ($\Delta \tau$) between primary and secondary transits of hot Jupiters and propose the most promising target for such detection, WASP-14 b. For this planet we expect a cumulated $\Delta \tau$ $\approx$ -250 s, due to tidal and relativistic precession, since its first photometric observations.
\end{abstract}

\begin{keywords}
celestial mechanics -- gravitation -- eclipses -- occultations -- planet-star interactions -- single targets: WASP-14 b
\end{keywords}



\section{Introduction}
\label{sec:intro}

Since the first exoplanet detection \citep{mayor1995}, a rich variety of planets orbiting solar-like stars has been discovered. Among this yield, transiting hot Jupiters (HJs) with short orbital periods and relatively high eccentricity offer the possibility to test General Relativity\footnote{For a comprehensive review on GR see \citet{debono2016}} (GR) through timing variations of transit observables caused by both GR-induced precession of their periastrons \citep{jordan2008, pal2008, iorio2011, iorio2011_2, iorio2016} and by oblateness \citep{iorio2011_2, iorio2016}. Given the relatively small entity of GR precession rates, a long time-span between the first and the last photometric observations of these planets is required to access the precision level needed by such measurement \citep{jordan2008}, which is now available for a considerable number of transiting HJs.

The major obstacle in determining the periastron precession is the high uncertainty usually associated with the value of the argument of periastron, which excludes the feasibility of measuring the precession rate from the inferred values of orbital parameters. However, a timing variation in the epoch of transits and occultations, or in the transit duration \citep[][I11 hereafter]{iorio2011}, can be used to detect GR precession rates if the observed planet has sufficiently short period and large orbital eccentricity. Since orbital precession is a cumulative effect, shorter periods lead to more orbits cumulated in the same time-span, hence HJ are inherently favoured. However, planets with longer periods are more likely to have large eccentricities, as we will discuss in Sec.~\ref{sec:precession}, and recently this motivated attempts \citep{blanchet2019} to detect GR precession also in planets with orbital period $\gtrsim$ 100 days.

\section{Precession effects on timing of transiting exoplanets}
\label{sec:precession}

A single exoplanet orbiting a solar-like star can be treated as a weak field approximation of Einstein's field equations of General Relativity \citep[GR,][]{miraldaesculde2002, pal2008}.
Their solution for closed orbits leads to a prograde precession of the argument of periastron ($\omega$). Over an orbital period $P$, the precession angle $\Delta \omega_P$ is given by \citep{weinberg1972, misner1973}:
\begin{equation}
\label{eq:prec}
    \Delta \omega_P = \dfrac{6 \pi}{a(1-e^2)} \dfrac{GM}{c^2},
\end{equation}
where $e$ is the orbital eccentricity,
$a$ is the semi-major axis of the planetary orbit and $M$ is the total mass of the planetary system, given by the sum of the mass of the star $M_{\star} $ and the mass of the planet $M_{\mathrm{p}}$,
\begin{equation}
    M= M_{\star} + M_\mathrm{p}.
\end{equation}
On short time scales ($\sim$ 10 yr) and for a two-body 
system we consider the orbital parameters as constant quantities, so that also $\Delta \omega_P$ and $P$ are constant. In this case $\omega$ varies linearly with time, because its time derivative
\begin{equation}
\label{eq:omega_dot}
    \Dot{\omega} = \dfrac{\Delta \omega_P}{P}
\end{equation}
does not change. Integrating Eq.~\ref{eq:omega_dot} over time we get
\begin{equation}
    \omega (t) = \omega_0 + \Dot{\omega} (t - t_0),
\end{equation}
where $t_0$ is a reference time at which $\omega (t_0) = \omega_0$. The time scale of GR precession is typically extremely long. Let us consider a slightly eccentric ($e$ = 0.1) planet in a close-in ($P$ = 2 days) orbit around a Sun-like star ($M_{\star}$ = 1 $M_{\odot}$): the GR precession rate amounts to 
$\sim 3.5 \times 10^{-11}$~rad
per orbit, which means that a complete precession requires $\sim 1$~Gyr.
From the Kepler's third law
\begin{equation}
\label{eq:kepler3}
    a^3 n^2 = GM,
\end{equation}
where $n = 2 \pi / P$ is the mean motion of the planet, we can write the semi-major axis of the planetary orbit as
\begin{equation}
\label{eq:new_a}
    a = \left( \dfrac{GM}{n^2}\right)^{1/3}.
\end{equation}
Substituting Eq.~\ref{eq:new_a} in Eq.~\ref{eq:prec} leads to
\begin{equation}
\label{eq:omega_dot2}
    \Dot{\omega} = \dfrac{3 n^{5/3}}{(1-e^2) c^2} (GM)^{2/3}.
\end{equation}
We used the Kepler's third law under the assumption that the time scale of our observational time-span is smaller than the time scale of variations in the values of the orbital parameters due to classical effects. Among such variations, eccentricity damping is known to be the most effective mechanism acting on short period planets \citep{rasio1996}, but over $\sim 10$~yr the effects of circularization are negligible, since its time scale is \citep{husnoo2011, goldreich1966}
\begin{equation}
    \dfrac{\Dot{e}}{e} = - \dfrac{P}{21 \pi} \left( \dfrac{Q_\mathrm{p}}{k_{2,\mathrm{p}}} \right) \left( \dfrac{M_\mathrm{p}}{M_\mathrm{J}} \right) \left( \dfrac{a}{R_\mathrm{p}} \right)^5 \approx 5 \times 10^7 
    \mathrm{yr}
\end{equation}
for a typical HJ, where $Q_\mathrm{p}$
is the dissipation function and $k_{2, \mathrm{p}}$ is the Love number of the planet. Another possible source of variations for the orbital period is the Applegate effect \citep{applegate1992}, due to changes in the quadrupole moment of a magnetically active parent star. However, \cite{watson2010} estimated a variation ranging from $\sim$ 0.1 s to a few seconds, with an 11 years modulation, for a typical HJ. Given the orders of magnitude of both the expected precession rate and its uncertainty, that we will discuss in Sec.~\ref{sec:wasp_14}, we can safely ignore the Applegate effect in the computation of our estimates.  
The Love number is a key parameter of the planetary tidal response to an outer gravitational potential and it is the main one in the computation of precession due to Newtonian sources \citep{SSD, cowling1938}.
Eq.~\ref{eq:omega_dot2} expresses the rate of precession as a function of the stellar and planetary masses. A measure of $\Dot{\omega}$ can be used both to test GR with exoplanets and to probe the mass of the star and the planet without relying on stellar models. However, measuring the argument of periastron with the accuracy and precision required is a daunting task. A small eccentricity ($e < 0.1$) means that the Keplerian model is highly degenerate in $\omega$, and unfortunately this is the common case with close-in planets, because of the eccentricity damping due to tidal forces. Radial Velocity (RV) measurements alone typically do not allow us to pin down $\omega$ with the required precision and accuracy. On the other hand, GR effects are best observable for close-in orbits. In this work we show how to achieve a good trade-off between these two needs, by addressing the effects of precession on a particular transit observable and by proposing an appropriate target.

\subsection{Effects of precession on the interval between primary and secondary transits}

\cite{pal2008} and I11 derived the equations for the effect of GR and classical precession on the observables of the transit method GR precession directly affects the argument of periastron $\omega$. In our approach we considered how transit observables are sensible to variations of $\omega$. Usually the values of $\omega$ from orbital solutions derived modelling the photometry of transits have large uncertainties. The observations of secondary eclipses (or occultations) allow us to remove the degeneracy in $\omega$, which can be fitted within a narrow range of probable values. The formal expression for the time interval $\tau$ between the mid-transit time of the primary and the secondary eclipses has been derived in the context of binary stars \citep{dong2013} and it is given by
\begin{equation}
\label{eq:tau}
    \tau = \dfrac{P}{\pi} \left[ \arccos{ \left( \dfrac{e \cos \omega}{\sqrt{1-e^2 \sin^2 \omega}} \right) } - \dfrac{e \sqrt{1-e^2} \cos{\omega}}{1- e^2 \sin^2 \omega} \right],
\end{equation}
which is accurate up to a term proportional to $e \cot^2 i$ \citep{jordan2008, sterne1940}. For a few degrees deviation from $i$ = 90$^{\circ}$ and small eccentricities ($e \lesssim$ 0.1), the additional term to Eq.~\ref{eq:tau} is smaller than $\sim$10$^{-3}$ $\tau$ \citep{sterne1940} and does not vary with time. So in this work it will be neglected.
\begin{figure}

	\includegraphics[width=\columnwidth]{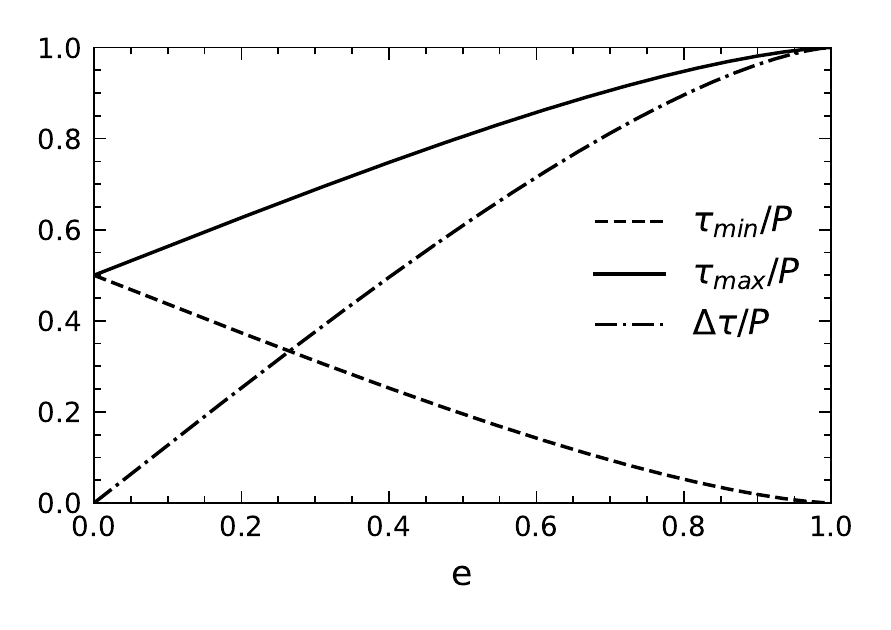}
    \caption{Variation of $\tau$ (dashed-dot line) in units of the orbital period for different values of the eccentricity. The time interval between the transit and the occultation of the planet is exactly half of the orbital period for circular orbits. As the eccentricity grows, $\tau$ oscillates between a minimum (dashed line) and a maximum value (solid line) due to the degeneracy in $\omega$.}
    \label{fig:e_vs_delta_tau}

\end{figure}
\begin{figure}

	\includegraphics[width=\columnwidth]{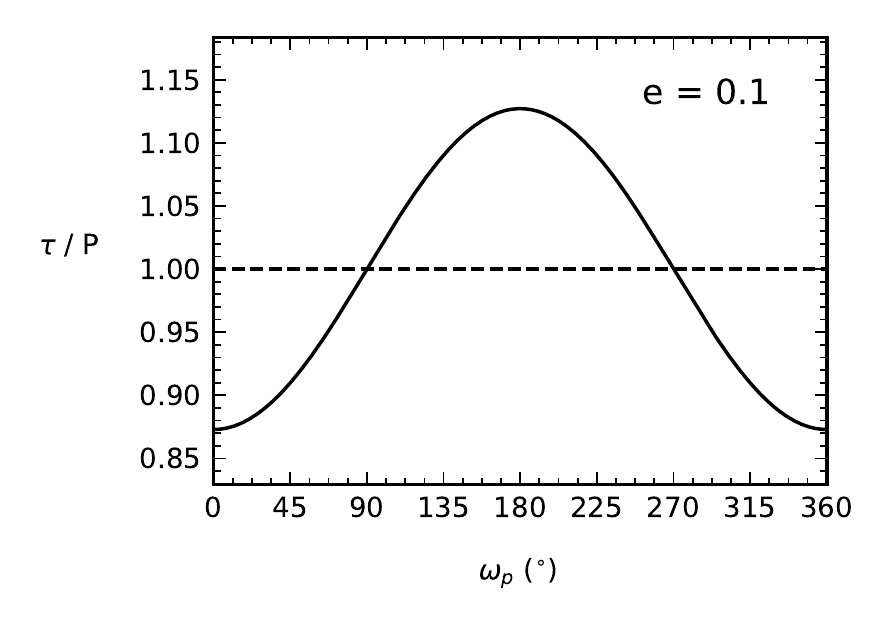}
    \caption{$\tau$ is a function of the argument of planetary periastron $\omega$, here computed using $P$ = 2 $\mathrm{d}$ and $e$ = 0.1. $\tau$ has a maximum for $\omega$ = 180$^\circ$ and a minimum around 0$^\circ$. The method presented in this work relies on the variation of $\tau$ with time, that is with $\omega$ = $\omega(t)$, so the most favourable cases are the ones where the time derivative of $\tau$ is larger (planets with $\omega$ around 90$^\circ$ and 270$^\circ$).}
    \label{fig:omega_vs_tau}

\end{figure}
As we can see in Fig.~\ref{fig:e_vs_delta_tau} and Fig.~\ref{fig:omega_vs_tau}, $\tau$/$P$ grows monotonically with the orbit eccentricity and has a maximum for $\omega$ = 180$^\circ$. Thus the feasibility of a measurement of $\tau$ is determined not only by the shape of the orbit of the planet, but also by its orientation in space with respect to the line of sight of the observer. Such considerations play a fundamental role in the choice of an appropriate target for the methods we are proposing in this work and will be discussed in Sec.~\ref{sec:wasp_14}. 
\par
However, before assessing the details of the $\tau$ measurements, we shall expose the link between the precession rates expressed as time variation of $\omega$, and the time variation of $\tau$.
To see how $\Dot{\omega}$ and $\tau$ are linked, let us consider Eq.~\ref{eq:tau}. The second factor on the right-hand side of this equation is a function of $e$ and $\omega$ only, and we shall call it $g(e, \omega )$, so that
\begin{equation}
\label{eq:tau_g}
    \tau = \dfrac{P}{ \pi } g ( e, \omega ) .
\end{equation}
As we mentioned in Sec.~\ref{sec:intro}, for small time scales $P$ and $e$ can be considered constant, while $\omega$ varies linearly. If we consider two reference times, $t_1$ and $t_2$, the variation of $\tau$ during the time-span of photometric observations $\Delta t =  t_2 - t_1$ (with $t_2$ $>$ $t_1$) is
\begin{equation}
\label{eq:delta_tau}
    \Delta \tau =  \dfrac{P}{\pi} [g(e, \omega(t_2)) - g(e, \omega(t_1))].
\end{equation}
Measuring such quantity is a good choice to asses the importance of variations in $\omega$ due to precession. It involves a differential measurement and not an absolute one, and it does not depend upon the extent of the observational time-span, as it would be if we wanted to measure a cumulated variation of the mid-transit ephemeris. With our choice of observables, the uncertainty in the measurements only depends upon the uncertainties on the transit and secondary eclipses photometry at the time they are observed.

Since we are interested in the $\tau$ dependency of Eq.~\ref{eq:omega_dot2}, let us take the time derivative of Eq.~\ref{eq:tau_g},
\begin{equation}
\label{eq:tau_dot}
\Dot{\tau} = \dfrac{P}{\pi} \Dot{g} = \dfrac{P}{\pi} \dfrac{\partial g}{\partial \omega} \Dot{\omega},
\end{equation}
where we assumed both $e$ and $P$ constant. Eq.~\ref{eq:tau_dot} can be reversed into
\begin{equation}
\label{eq:new_omega_dot}
\Dot{\omega} = \dfrac{\pi}{P} \left( \dfrac{\partial g}{\partial \omega} \right)^{-1} \Dot{\tau}.
\end{equation}
The partial derivative of $g(e, \omega)$ with respect to $\omega$ is quite intricate, but it can be easily expressed if we expand $\partial g / \partial \omega$ in powers of $e$. To the third order we have
\begin{equation}
\label{eq:dg_domega}
\dfrac{\partial g}{\partial \omega} = 2 e \sin \omega + \mathcal{O}(e^3),
\end{equation}
and consequently
\begin{equation}
\label{eq:tau_dot_approx}
\Dot{\tau} = \dfrac{P}{\pi} 2 e \sin \omega \: \Dot{\omega}.
\end{equation}
In our scenario, $\Dot{\omega}$ and $\Dot{\tau}$ are constant quantities, so we can derive the expected variation of $\tau$ simply integrating Eq.~\ref{eq:tau_dot_approx} over the observational time-span $\Delta t$,
\begin{equation}
\label{eq:delta_tau_definetion}
    \Delta \tau = \dfrac{2}{\pi} \: P \: \Dot{\omega} \: \Delta t \: e \sin \omega.
\end{equation}

\subsection{GR precession rate}
\label{sec:GR_effects}

In order to evaluate the $\Delta \tau$ variation we need to compare theoretical predictions from GR to the photometric observations, so we need to express the $\Dot{\omega}$ of Eq.~\ref{eq:omega_dot2} as a function of $\Dot{\tau}$, which can be easily done using Eq.~\ref{eq:new_omega_dot} and its approximation given by the expansion that led us to Eq.~\ref{eq:tau_dot_approx}. Therefore, we begin by substituting Eq.~\ref{eq:new_omega_dot} into Eq.~\ref{eq:omega_dot2}. We obtain
\begin{equation}
\label{eq:gm_implicit}
GM = \dfrac{c^3}{12 \pi \sqrt{6}} (1-e^2)^{3/2} \left[ \dfrac{\Dot{\tau}}{\partial g / \partial \omega} \right]^{3/2} P,
\end{equation}
which allows us to write a second order approximated form of Eq.~\ref{eq:gm_implicit} as
\begin{equation}
\label{eq:gm_final}
M = \dfrac{c^3}{48 \pi \sqrt{3} G} \left( \dfrac{1 - e^2}{e} \right)^{3/2} \left( \dfrac{\Dot{\tau}}{\sin \omega} \right)^{3/2} P.
\end{equation}
Eq.~\ref{eq:gm_final} gives the mass of the planetary system as a function of both primary and secondary transit observables. As we assumed, during a time-span $\Delta t$ $\sim$ 10 yr, the time derivative $\Dot{\tau}$ is constant and therefore can be computed using
\begin{equation}
\label{eq:delta_tau_over_delta_t}
\Dot{\tau} = \dfrac{\Delta \tau}{\Delta t}.
\end{equation}
In this case Eq.~\ref{eq:gm_final} is equivalent, up to a second order approximation in $e$, to
\begin{equation}
\label{eq:m_approx}
M = 0.176 \mathrm{M}_{\odot} \times \left( \dfrac{P}{\mathrm{d}} \right) \left( \dfrac{\Delta \tau}{\mathrm{min}} \right)^{3/2} \left( \dfrac{\Delta t}{\mathrm{yr}} \right)^{-3/2} (e \sin \omega)^{-3/2}. 
\end{equation}
If we reverse Eq.~\ref{eq:gm_final} to express, again to the second order in orbital eccentricity, the expected GR variation of $\tau$, in seconds, given the mass of the planetary system, we get
\begin{equation}
\label{eq:delta_tau_GR}
\Delta \tau_\mathrm{GR} = 191 \mathrm{s} \times \left( \dfrac{M}{\mathrm{M}_{\odot}} \right)^{2/3} \left( \dfrac{P}{\mathrm{d}} \right)^{-2/3} \left( \dfrac{\Delta t}{\mathrm{yr}} \right) e \sin \omega. 
\end{equation}

\subsection{Tidal and rotational precession rates}
\label{sec:dynamical_effects}

I11 considered among the sources of precession: 1) the effects of tidal bulges, both on the planet and on the star; 2) the effects of planetary and stellar oblateness (non-zero $J_2$ coefficients); 3) the relativistic effect of stellar angular momentum (the Lense-Thirring effect); 4) the presence of another orbiting body. In I11 scenario, i. e. a Jupiter-mass planet orbiting a Sun-like star with $e = 0.07$ and $a = 0.04$ au, the GR effects and the tidal forces acting on the planet are the major factors contributing to the total orbital precession. All other effects are at least one order of magnitude smaller (see Eq. 72 in I11), with the possible exception of a third body orbiting the barycentre of the system. The effect of the latter factor strongly depends on the orbital configuration and the mass of the third body and it can vary by orders of magnitudes.

The Lense-Thirring precession has been estimated by I11 to produce an effect on $\Delta \tau$ 4 orders of magnitude smaller than the GR effect. As we will show in Sec.~\ref{sec:wasp_14}, the GR effect on $\Delta \tau$ for the HJ that we propose as an appropriate target is of the order of the hundreds of seconds. Consequently, according to I11, the Lense-Thirring effect in our case would account for a variation of just a few tens of milliseconds. Since the uncertainties on the values of orbital and physical parameters of the planetary system result in an uncertainty of $\approx$50 s on the final $\Delta \tau$ estimate, we decided to ignore the Lense-Thirring effect in the computation of our estimate of $\Delta \tau$.

Stellar oblateness effect on $\Delta \tau$, that is the effect of the quadrupole moment due to the non-zero $J_2$ coefficient of the star, is at most 2 orders of magnitude smaller than the GR effect. For our target, this translates into a few second variation of $\Delta \tau$. Moreover, the $\Delta \tau$ variation induced by such effect is proportional to
\begin{equation}
3 + 5 \cos 2 \Psi_{\star}
\end{equation}
(eq. 65 of I11), where $\Psi_{\star}$ is the inclination of the orbit with respect to the equatorial plane of the star, so the magnitude of this effect has a maximum for $\Psi_{\star}$ = 0$^{\circ}$ and a minimum for $\Psi_{\star}$ = 90$^{\circ}$. For our proposed target, $\Psi_{\star}$ = 33$^{\circ}$ $\pm$ 7$^{\circ}$ \citep{johnson2009}, so the stellar quadrupole effect is $\sim$60$\%$ of its maximum value, and it likely accounts for a $\sim$1 s variation of $\Delta \tau$. As for the Lense-Thirring precession, the uncertainties in the case of our target put the stellar quadrupole effect of $\Delta \tau$ beyond the threshold for detection, and therefore it has been excluded from our computations.

Tidal forces are symmetrical to the orbital plane and they raise a bulge on the surface of planets. The only non-vanishing component is the radial one \citep{SSD} and perturbation theory for close-in binaries \citep{cowling1938} gives the expression for the associated acceleration along the radial direction,
\begin{equation}
\label{eq:ar_tidp}
a^{\mathrm{tid,p}}_\mathrm{R} = - \left(  \dfrac{M_{\star}}{M_\mathrm{p}} \right) \dfrac{3 k_{2 \mathrm{p}} R_\mathrm{p}^5 GM}{r^7}.
\end{equation}
Such radial acceleration depends on how the planetary interiors react to the potential and they are the source of the tidal bulge on the planetary surface. The information about the internal structure of the planet is encapsulated in the parameter $k_{2\mathrm{p}}$, the second Love number, which is the ratio between the second order of the potential induced by the tidal deformation \citep{ragozzine2009} and the second order of the external potential applied to the planet ($V_{2}^{\mathrm{app}}$). So if we have 
\begin{equation}
    k_{2,\mathrm{p}} \equiv \frac{V_{2}^{\mathrm{ind}}}{V_{2}^{\mathrm{app}}}
\end{equation}
then, following \citet{SSD} and \citet{ragozzine2009},
we can express $k_{2,\mathrm{p}}$ as a function of the $J_{2, \mathrm{p}}$ moment as
\begin{equation}
\label{eq:j2}
k_{2, \mathrm{p}} V_{2}^{\mathrm{app}} = - J_{2, \mathrm{p}} \dfrac{G M_\mathrm{p}}{R_\mathrm{p}} P_2 (\cos \theta),
\end{equation}
where $\theta$ is the planetary colatitude and $P_2$ is the second order Legendre polynomial. The applied potential ($V^{\mathrm{app}}_{2}$) has a centrifugal component due to planetary rotation,
\begin{equation}
\label{eq:V_2_r}
    V_{2, \mathrm{r}} = \dfrac{\Omega^2 R_\mathrm{p}^3}{G M_\mathrm{p}},
\end{equation}
where $\Omega$ is the rotational angular frequency, and a tidal component due to the stellar gravity
\begin{equation}
\label{eq:V_2_t}
    V_{2, \mathrm{t}} = -3 \left( \dfrac{R_\mathrm{p}}{r} \right)^3 \left( \dfrac{M_{\star}}{M_\mathrm{p}} \right). 
\end{equation}
If there is no obliquity, which means $\theta$ = 0 and consequently $P_2 (\cos \theta)$ = -1/2, then Eq.~\ref{eq:j2} can be rearranged to give a convenient expression for $J_{2, \mathrm{p}}$,
\begin{equation}
    J_{2, \mathrm{p}} = \dfrac{k_{2, \mathrm{p}}}{3} ( V_{2, \mathrm{r}} - \dfrac{V_{2, \mathrm{t}}}{2} ).
\end{equation}
Since $V_{2, \mathrm{t}}$ is a function of the relative distance between the star and the planet and consequently of time, then also $J_{2, \mathrm{p}}$ for eccentric planets is a function of time. Short-period massive planets are likely tidally locked \citep{guillot1996, SSD}, so in Eq.~\ref{eq:V_2_r} the angular frequency of the planetary rotation is equal to its mean motion,
\begin{equation}
\label{eq:Omega_omega}
    \Omega = n,
\end{equation}
and then, by Kepler's Third Law, as expressed in Eq.~\ref{eq:kepler3}, we obtain
\begin{equation}
\label{eq:V_2_app_tidally_locked}
    V_2^{\mathrm{app}} = \left( \dfrac{R_{\mathrm{p}}}{a} \right)^3 \left( \dfrac{M_{\star} + M_{\mathrm{p}}}{M_{\star}} \right).
\end{equation}
Since $M_{\star}$ $\gg$ $M_{\mathrm{p}}$, we can approximate Eq.~\ref{eq:V_2_app_tidally_locked} to
\begin{equation}
    V_{2, \mathrm{r}} \approx -3 V_{2, \mathrm{t}},
\end{equation}
which along with Eq.~\ref{eq:j2} gives
\begin{equation}
    J_{2, \mathrm{p}} = \dfrac{5}{6} k_{2, \mathrm{p}} \left( \dfrac{M_{\star}}{M_\mathrm{p}} \right) \left( \dfrac{R_\mathrm{p}}{a} \right)^3.
\end{equation}
This is a simple relation between two parameters, $J_{2, \mathrm{p}}$ and $k_{2, \mathrm{p}}$, that hides the complex behaviour of planetary interiors subject to external potentials. The second Love number would be zero if all the mass was concentrated in the core, and solar-like stars, that have relatively massive cores, are expected to have $k_{2, \mathrm{p}}$ $\sim$ 0.03 \citep{claret1995}.
A fairly good model for a cold giant planet is a polytrope with $n$ = 1 and it would have $k_{2, \mathrm{p}} = 0.52$ \citep{kopal1959}, while a uniform sphere would have $k_{2, \mathrm{p}} = 3/2$ \citep{SSD}. In fact, Jupiter and Saturn show respectively $k_{2, \mathrm{p}} = 0.49$ and $0.32$, marking with a $\sim$ 0.1 the difference between a less and a more massive core.
Now that we have defined the role of the internal structure in tidal and rotational potentials, we can use results from binary stars theory \citep{sterne1939} to write the tidal precession rate,
\begin{equation}
\label{eq: omega_dot_tid}
    \Dot{\omega}_{\mathrm{tid,p}} = \dfrac{15}{2} k_{2, \mathrm{p}}  \left( \dfrac{R_\mathrm{p}}{a} \right)^5 \dfrac{M_{\star}}{M_\mathrm{p}} n f_2 (e),
\end{equation}
and the rotational precession rate,
\begin{equation}
\label{eq: omega_dot_rot}
    \Dot{\omega}_{\mathrm{rot,p}} = \dfrac{1}{2} k_{2, \mathrm{p}}  \left( \dfrac{R_\mathrm{p}}{a} \right)^5 \dfrac{\Omega^2 a^3}{G M_\mathrm{p}} n g_2 (e),
\end{equation}
where $f_2 (e)$ and $g_2 (e)$ are polynomial functions of the orbital eccentricity, as function of stellar and orbital parameters. If we use the hypothesis of tidally locked planets that led to Eq.~\ref{eq:Omega_omega}, in the above equations, we note that $\Dot{\omega}_\mathrm{{rot,p}}$ and $\Dot{\omega}_\mathrm{{tid,p}}$ differ only by a factor proportional to the ratio
\begin{equation}
   \dfrac{g_2 (e)}{f_2 (e)} \approx 1 - \dfrac{3}{2} e^2,
\end{equation}
at the second order approximation in $e$. We can combine Eq.~\ref{eq: omega_dot_rot} and Eq.~\ref{eq: omega_dot_tid} into a single definition of a dynamical precession rate,
\begin{equation}
    \Dot{\omega}_\mathrm{{dyn,p}} = \Dot{\omega}_\mathrm{{tid,p}} +  \Dot{\omega}_\mathrm{{rot,p}} = \left( 1 + \dfrac{1}{15} \dfrac{g_2 (e)}{f_2 (e)} \right) \Dot{\omega}_\mathrm{{tid,p}}.
\end{equation}
To the second order in eccentricity, such dynamical precession rates can be written as
\begin{equation}
    \Dot{\omega}_\mathrm{{dyn,p}} = \left( \dfrac{16}{15} - \dfrac{1}{10} e^2 \right) \Dot{\omega}_\mathrm{{tid, p}}.
\end{equation}
Substituting the precession rate $\Dot{\omega}_\mathrm{dyn,p}$ in Eq.~\ref{eq:tau_dot_approx} we get the expected value of $\Dot{\tau}$ due to planetary rotation and tidal forces,
\begin{equation}
    \Dot{\tau}_\mathrm{{dyn}} = \dfrac{32}{15 \pi} e \sin \omega \: P \: \Dot{\omega}_\mathrm{{tid,p}} + \mathcal{O}(e^2),
\end{equation}
and as a function of orbit and planet parameters
\begin{equation}
    \Dot{\tau}_\mathrm{{dyn}} = 32 e \sin \omega \: k_{2, \mathrm{p}} \: \left( \dfrac{R_\mathrm{p}}{a} \right)^5 \dfrac{M_{\star}}{M_\mathrm{p}}.
\end{equation}
Finally, the time-integrated value $\Delta \tau$, as given by Eq.~\ref{eq:delta_tau_definetion}, is
\begin{equation}
\label{eq:delta_tau_dyn}
    \Delta \tau_\mathrm{{dyn}} = 1.01 \cdot 10^9 \mathrm{s} \times \left( \dfrac{R_\mathrm{p}}{a} \right)^5 \left( \dfrac{M_{\star}}{M_\mathrm{p}} \right) \left( \dfrac{\Delta t}{\mathrm{yr}} \right) e \sin \omega \: k_{2,\mathrm{p}}.
\end{equation}
Combining Eq.~\ref{eq:delta_tau_definetion} and Eq.~\ref{eq:delta_tau_dyn} we obtain the total effect of the precessione rate on the time-integrated value:
\begin{equation}
\label{eq:delta_tau_tot}
    \Delta \tau_{\mathrm{tot}} =  \Delta \tau_{\mathrm{GR}} + \Delta \tau_{\mathrm{dyn}}.
\end{equation}
Finally, we have obtained in Eq.~\ref{eq:delta_tau_tot} a quantity, $\Delta \tau_{\mathrm{tot}}$, which is a function of GR precession rate and it is the observable that astronomers can measure from the photometry of planetary transit and secondary eclipses.


\section{The case of WASP-14 b}
\label{sec:wasp_14}

\begin{table*}
\centering
 \begin{tabular}{l c c c c c} 
 \hline
 Parameter & Symbol (unit) & Value from \cite{bonomo2017} & Value from \cite{joshi2009} & Value from \cite{wong2015}\\ [0.5ex] 
 \hline\hline
 Orbital period & $P$ (day) & 2.2437661 $\pm$ 1.1x$10^{-6}$ & 2.243752 $\pm$ 1.0x$10^{-5}$ & 2.24376524 $\pm$ 4.4x$10^{-7}$ \\
 Orbital semi-major axis & $a$ (au) & 0.0358$^{+0.0013}_{-0.0012}$ & 0.036 $\pm$ 0.001 & 0.0371 $\pm$ 0.0011 \\
 Orbital eccentricity & $e$ & 0.0782$^{+0.0012}_{-0.0014}$ & 0.091 $\pm$ 0.003 & 0.0830$^{+0.0030}_{-0.0029}$ \\
 Orbital inclination & $i$ ($^{\circ}$) & --- & 84.32$^{+0.67}_{-0.57}$ & 84.63 $\pm$ 0.24 \\
 Argument of periastron & $\omega$ ($^{\circ}$) & 251.61$\pm$ 0.41 & 253.371$^{+0.693}_{-0.678}$ & 252.67$^{+0.77}_{-0.70}$\\
 Planet mass & $M_{\mathrm{P}}$ (M$_{\mathrm{Jup}}$) & 7.22$^{+0.50}_{-0.49}$ & 7.34$^{+0.51}_{-0.50}$ & 7.76 $\pm$ 0.47 \\
 Planet radius & $R_{\mathrm{P}}$ (R$_{\mathrm{Jup}}$) & 14.36$^{+0.84}_{-0.92}$ & 14.36$^{+0.84}_{-0.92}$ & 13.69 $\pm$ 0.46\\
 Stellar mass & $M_{\star}$ (M$_{\odot}$) & 1.21$^{+0.12}_{-0.13}$ & 1.21$^{+0.13}_{-0.12}$ & --- \\
 Stellar radius & $R_{\star}$ (R$_{\odot}$) & 1.306$^{+0.066}_{-0.073}$ & --- & 1.306 $\pm$ 0.073\\
 Transit duration & $T_{14}$ (hour) & --- & 3.06$^{+0.0672}_{-0.0744}$ & --- \\
 [1ex] 
 \hline
 \end{tabular}
\caption{A summary of the orbital and physical properties of the WASP-14 system, from its discovery paper \citep{joshi2009}, from the most recent homogeneous analysis of transiting giant planets \citep{bonomo2017}, and from \citet{wong2015}, where the authors analyzed the thermal phase curves and the occultations of the planet and which we used as reference in this work, with the exception of the stellar mass. The stellar mass of WASP-14 has been taken from \citet{southworth2012}, a homogeneous study of 38 transiting planets where the author reviewed the limb darkening coefficients taking into account the effects of non-zero orbital eccentricity and possible contamination light.}
\label{table:1}
\end{table*}

From what we demonstrated in the previous Section, an ideal case to test the effects of GR precession is a massive transiting exoplanet in a close-in, eccentric orbit. While hundreds of HJs have been discovered \citep{dawson2018}, most of them have non-detectable or very small eccentricities, due to a well-known circularization process driven by tidal forces, as mentioned in Section~\ref{sec:dynamical_effects}; only a small fraction of HJs have $e>0.05$. An additional constraint is that the host star has to be bright enough to enable high-precision follow-up observations. This is particularly true for the timing of the secondary eclipse (usually carried out in the near infrared region), which in turn should be deep enough to mitigate the impact of systematic errors on the photometric measurements.

\begin{figure}
	\includegraphics[width=\columnwidth]{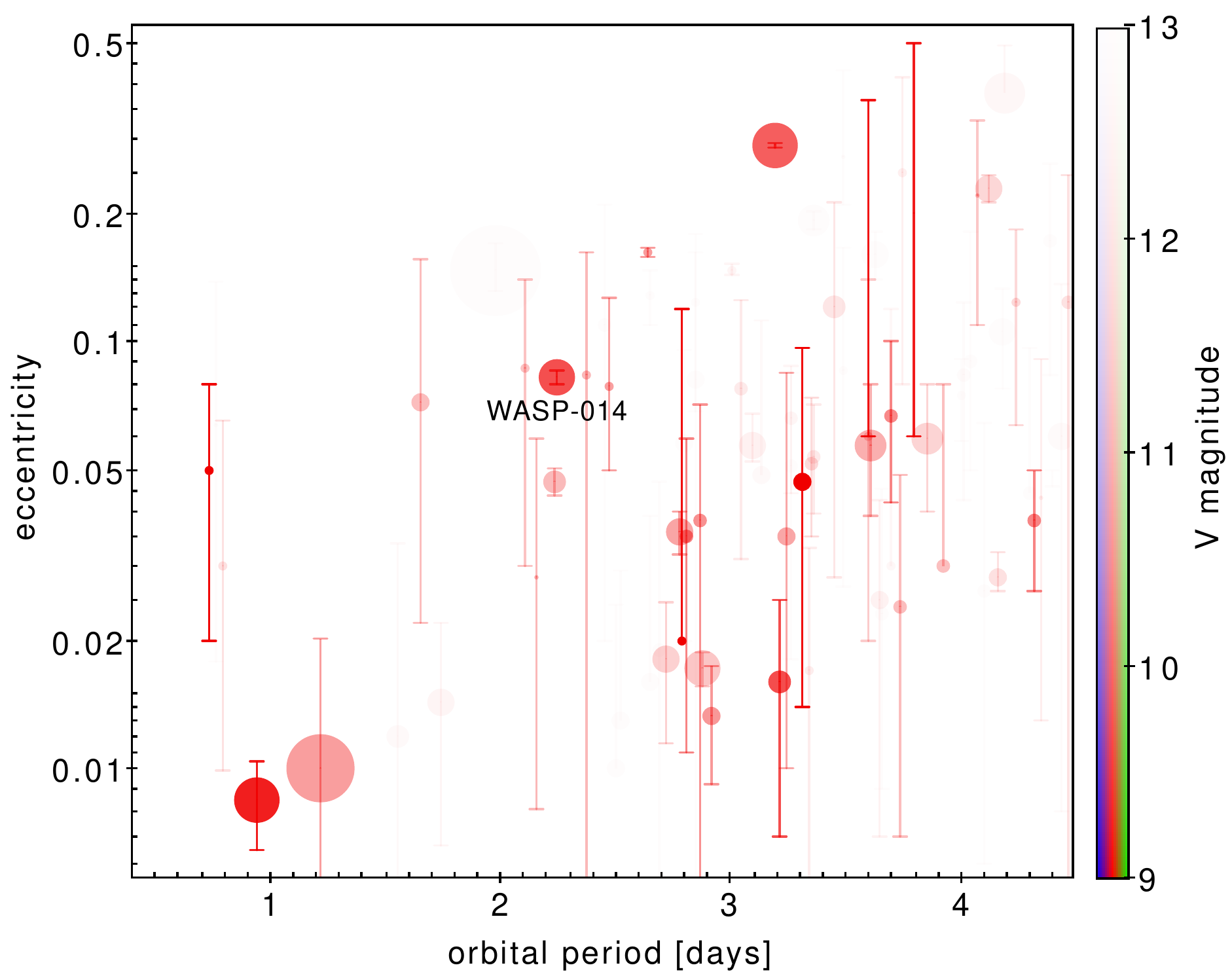}
    \caption{Known transiting planets (source: TEPCAT on 2020/10/01, \citealt{southworth2011}) as a function of their orbital period and eccentricity; the V magnitude of the host star is color-coded as a transparency mask, while the planetary mass is proportional to the area of each point. WASP-14 b (labeled) clearly stands out as one of the most eccentric HJs hosted by a bright star.}
    \label{fig:w14}
\end{figure}

\begin{figure}
    \centering
    \includegraphics[width=\columnwidth]{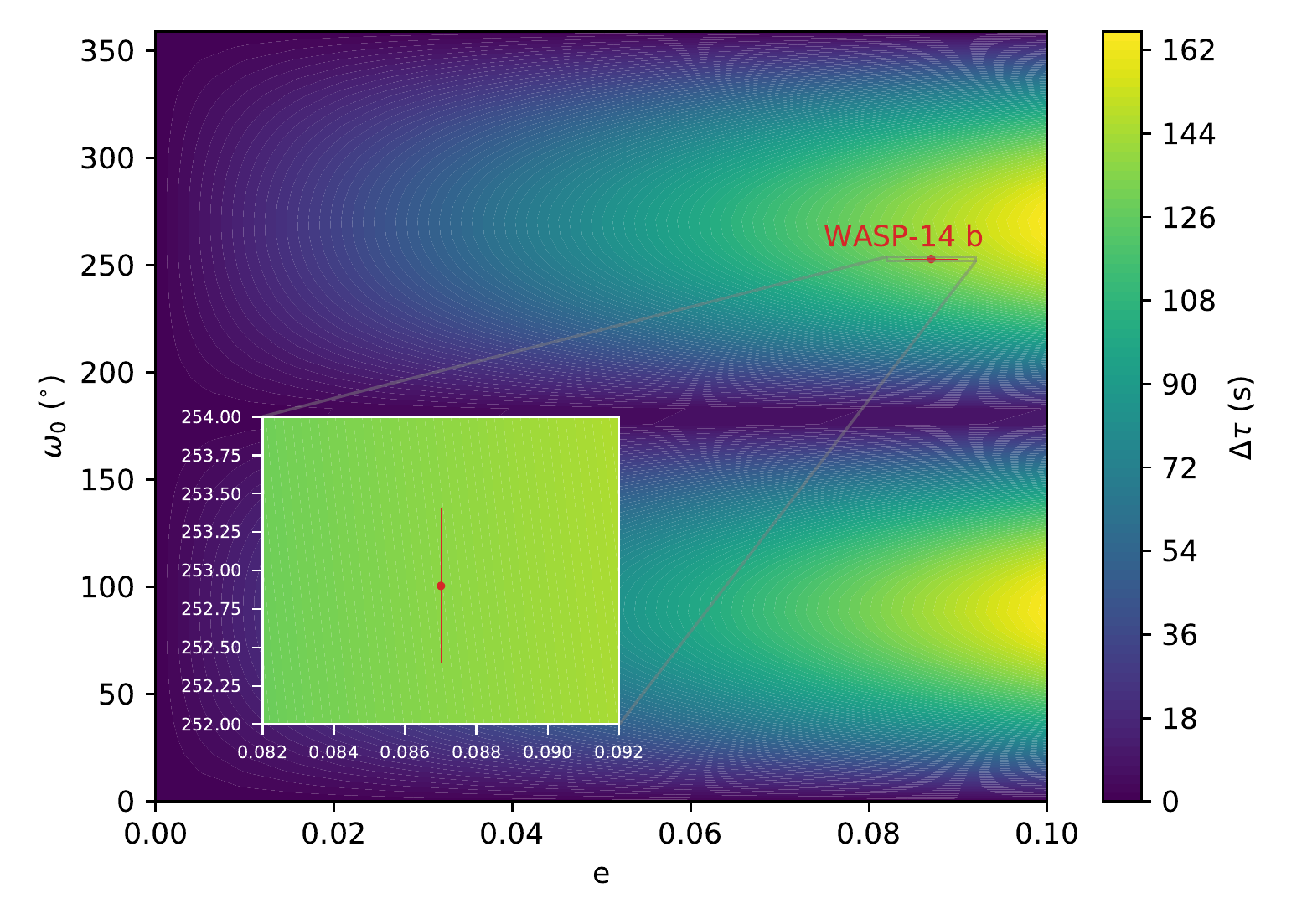}
    \caption{The expected $\Delta \tau$ variation due to GR for WASP-14 b, expressed as a function of orbit eccentricity and argument of periastron over a 12 year observational time-span is $\sim 126$~s.}
    \label{fig:map}
\end{figure}

When comparing the distribution in eccentricity, orbital period and magnitude of all the known HJs (Fig.~\ref{fig:w14}), and the planets with the deepest secondary eclipse and the highest expected $\Delta \tau$ (Tab.~\ref{table:2}), WASP-14 b clearly stands out as one of the most favorable targets for our study. Also, the orientation of the orbit of WASP-14 b seem to be favourable for a measurement of the GR $\Delta \tau$, as we can see from Fig.~\ref{fig:map}.

In what follows we will focus on WASP-14 b as a possible application to test our model.

WASP-14 b is a transiting, massive HJ \citep{joshi2009} with $M_{\mathrm{p}}$ = 7.76 $\pm$ 0.47 $\rm M_J$ \citep{wong2015}. Its orbit is slightly eccentric, with e = 0.0830$^{+0.0029}_{-0.0030}$ \citep{wong2015}, and close-in, since $P$ = 2.24376524 $\pm$ 4.4x$10^{-7}$~d \citep{wong2015} around a bright F5 star with $V=9.7$, $K=8.6$ and $M_{\star}$ = 1.35 $\pm$ 0.12 $\rm M_{\odot}$ \citep{southworth2012}. See Tab.~\ref{table:1} for a summary of the orbital and physical properties of the WASP-14 system.
WASP-14 b is a well-studied planet, both with transit photometry \citep{joshi2009} and with the radial velocity method \citep{husnoo2011, joshi2009}. Transit Time Variation (TTV) analysis showed no evidence for any other planet in the system \citep{raetz2015} and the observations of secondary eclipses made it possible to measure the orbital parameter values with great accuracy and precision \citep{blecic2013}. Among HJs with detected secondary eclipses \citep{garhart2020}, WASP-14 b has one the deepest ones, $\sim$ 1900~ppm at 3.6~$\mu$m and $\sim$ 2400~ppm at 4.5~$\mu$m \citep{wong2015}.  
Luckily, there is an extensive record of publicly available high-quality photometric observations of WASP-14 b relevant to our study, including:
\begin{enumerate}
    \item Two high-precision transit light curves at epoch 2008.0 \citep{joshi2009} and 2009.5 \citep{johnson2009}, gathered respectively with the RISE instrument in the R+V wide band and with the UH-2.2m telescope in the $r'$ band. These two observations, combined, set a first epoch for the transit time with an accuracy of about 20~s. Other more recent transit light curves from \citet{raetz2015} pinpoint the $T_0$ with a similar level of precision at epoch $\sim$2012.0;
    \item Three secondary transits gathered with Spitzer at 3.6, 4.5, 8 $\mu$m respectively \citep{blecic2013}, in 2009 and 2010. When combined, those observations constrain the timing of the occultation with a precision of $\sim$40~s around epoch 2010.0. Even better, the Spitzer phase curves published by \citet{wong2015} includes two full eclipses at 3.6~$\mu$m and two at 4.5~$\mu$m, which can be combined into a single point at epoch 2012.4 with $\sim$35~s of precision. The same data also include two primary transits (one in each filter) further constraining the transit timing at the $\sim$15~s level around the same epoch. 
\end{enumerate}

\begin{figure}

	\includegraphics[width=\columnwidth]{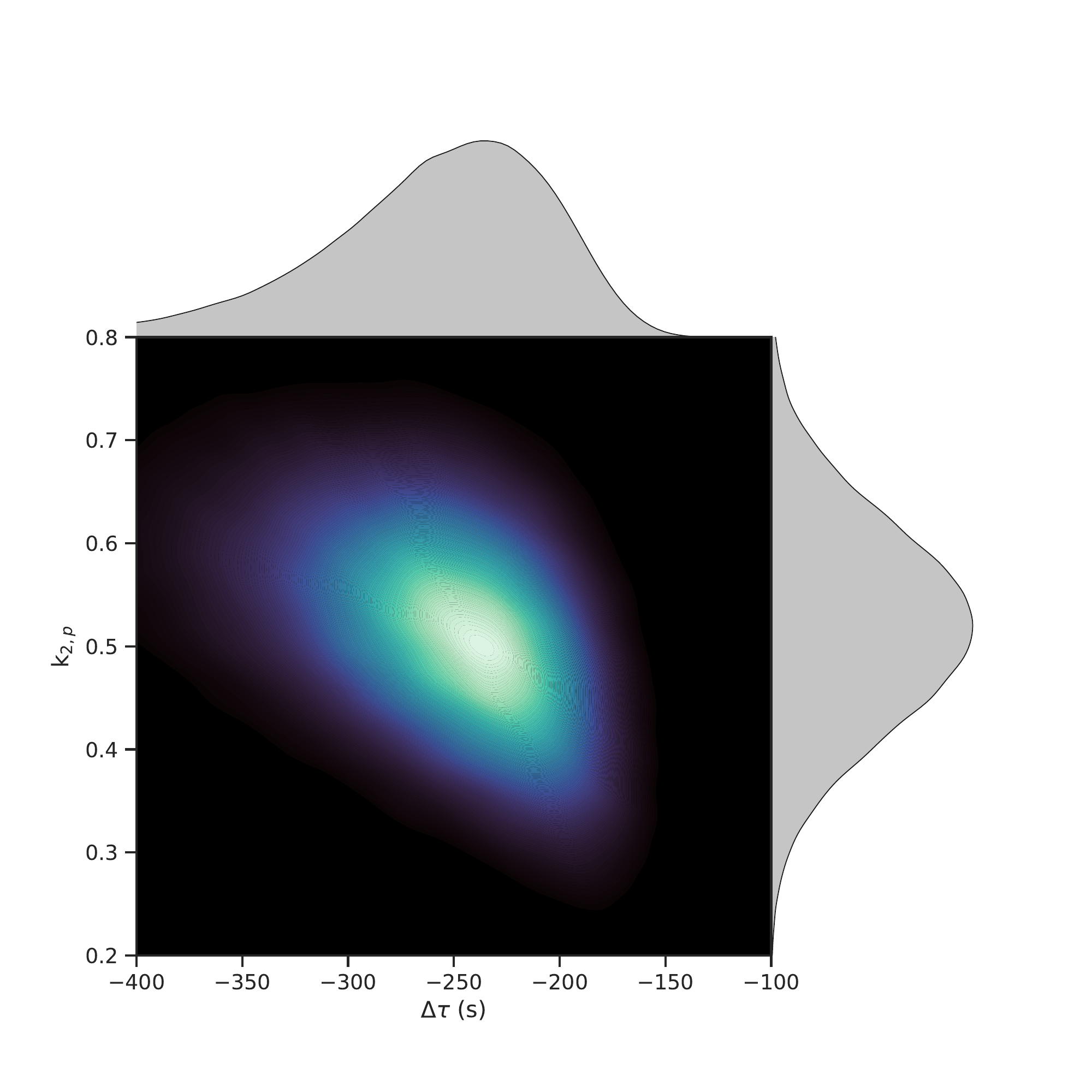}
    \caption{Joint probability density for $\Delta \tau$ and $k_{2,p}$ if the latter is assumed to follow a normal distribution with $\mu$ = 0.5 and $\sigma$ = 0.1 }
    \label{fig:k2_vs_delta_tau2}

\end{figure}

With these premises, WASP-14 b is then one of the best candidates for a detection of GR precession by measuring the variation $\Delta \tau$ as described in the previous sections. In order to get a realistic estimate of the expected $\Delta \tau$, we considered an observational time-span of 12 yr and the parameter values derived by \cite{joshi2009}, \cite{husnoo2011}, \cite{southworth2012}, \cite{blecic2013}, \cite{raetz2015}, and \cite{fontanive2019}. From Eq.~\ref{eq:delta_tau_GR} and Eq.~\ref{eq:delta_tau_dyn} we get
\begin{equation}
    \Delta \tau_\mathrm{{GR}} =  (-126.2 \pm 9.0) \: \mathrm{s}
\end{equation}
and
\begin{equation}
    \Delta \tau_\mathrm{{dyn}} =  (-262 \pm 97) \: \mathrm{s} \times k_{2, \mathrm{p}},
\end{equation}
where the uncertainties have been estimated with Monte Carlo (MC) simulations. The only unspecified parameter in the above equations is the planet Love number $k_{2, \mathrm{p}}$. We expect \citep{ragozzine2009, bodenheimer2001} a tidally locked HJ to have $k_{2, \mathrm{p}}$ between $\sim$0.1 and $\sim$0.6, so we draw samples from a uniform distribution with those limits,
\begin{equation}
    k_{2,\mathrm{p}} \sim \mathcal{U}(0.1, 0.6),
\end{equation}
and again via an MC simulation estimated an expected $\Delta \tau$ and its uncertainty,
\begin{equation}
\label{eq:final_result}
    \Delta \tau_\mathrm{{tot}} = (-218 \pm 53) \: \mathrm{s}.
\end{equation}
However, a more informative prior on $k_{2, \mathrm{p}}$ can be constructed if we assume that the most probable value for the Love number of an HJ is the same value a $n$ = 1 polytrope would require. As we mentioned in Sec.~\ref{sec:dynamical_effects}, \cite{kopal1959} derived $k_2$ = 0.52 for such a model, and in our solar system the difference between gaseous planets with relatively less or more massive cores is reflected by a $\sim$0.1 variation in the Love number value. With this information we conducted another MC simulation using as prior on $k_{2, \mathrm{p}}$ the normal distribution
\begin{equation}
    k_{2,\mathrm{p}} \sim \mathcal{N}(\mu = 0.5, \sigma = 0.1).
\end{equation}
In this scenario (see Fig.~\ref{fig:k2_vs_delta_tau2}), the expected value for $\Delta \tau$ is
\begin{equation}
    \Delta \tau_{\mathrm{tot}} = -253^{+63}_{-45} \: \mathrm{s},
\end{equation}

where we considered a 1-$\sigma$ interval centered on the median of the posterior distribution for $\Delta \tau_\mathrm{{tot}}$.

The stellar system of WASP-14 is known to host a star, WASP-14B, certainly associated with the main component \citep{ngo2015, fontanive2019} and another, more distant star, WASP-14C, whose membership of the system has been recently inferred from parallactic properties \citep{fontanive2019, gaiaDR2}. WASP-14B is an 0.33 $\pm$ 0.04 $\rm M_{\odot}$ stellar companion at a separation of 300 $\pm$ 20 AU \citep{ngo2015}, while WASP-14C is a K5 V star with $M_\mathrm{C}$ = 0.280 $\pm$ 0.016 $\rm M_{\odot}$ at a projected separation of $\sim$ 1900 AU \citep{fontanive2019}. Both the companions of WASP-14A are unlikely to have played any role in the formation and evolution of WASP-14 b \citep{fontanive2019}. Nonetheless they may, in principle, cause a slight variation in the precession rates from the values derived in this work. Assuming circular orbits about the barycenter of the system, WASP-14B has an orbital period of $\approx$4000 yr, and WASP-14C an orbital period of $\approx$52900 yr, which are, respectively, $\sim$10$^6$ and $\sim$10$^7$ times longer than the orbital period of the planet. During the 10 yr observational time-span their angular positions, as seen by the planet, have been almost fixed, since they are at most $\sim$8 arcminutes for WASP-14B and $\sim$4 arcminutes for WASP-14C. The effective impact of the two stellar companion of WASP-14A on the precession rate of WASP-14 b depends upon their gravitational potential at the planet position and it is a complex function of both the distance of the two stars from the planet and the orientations of the spins and the orbital planes of all the bodies involved \citep{iorio2011}. Such potentials are in the form \citep{hogg1991}
\begin{equation}
V_\mathrm{X} = \dfrac{GM_\mathrm{X}}{2 r_\mathrm{{X}}^3} \left[ r^2 - 3(\boldsymbol{r} \cdot \boldsymbol{l}_\mathrm{X}) \right],
\end{equation}
where X is the distant orbiting body (the B and C components of WASP-14 in our case), $r_\mathrm{{X}}$ is the distance between the planet and the distant orbiting body, $\boldsymbol{l}_\mathrm{X}$ is the vector pointing from the planet to the distant orbiting body and $r$ is the distance between the barycenter of the system and the planet. Since this potential scales with the third power of the distance of the two minor stellar companions of WASP-14A and that, during the observational time-span, such distance can be considered constant. Given the ~50$\%$ uncertainty on the predicted value of $\Delta \tau$, we can then safely ignore the perturbations of WASP-14B and C on the orbit of WASP-14 b. So the prediction of Eq.~\ref{eq:final_result} still holds valid.

Is such a timing drift ($\Delta\tau\sim 4$~min) detectable in the near future? We showed above that archival data of WASP-14 b constrain the timing of both the primary transit and occultation to better than 20~s and 40~s (respectively) at epoch 2009.0 and 2010.0, thus observations with a similar or better precision at any time of the future will be able to detect the drift with a $>5$~$\sigma$ significance. As for the primary transit, the TESS satellite \citep{ricker2014} is going to observe WASP-14 for the first time during its Sector 50 (i.e., from March 26 to April 22, 2022); by examining the actual photometric performances of TESS on two already observed HJs with a similar orbital period, depth and magnitude (WASP-95b and WASP-111b) we estimate that the TESS light curve will deliver a timing error of $<20$~s at epoch 2022.3. Even without TESS, light curves gathered from medium-sized, ground-based facilities could be equally effective in constraining the second epoch, since timing errors of $< 30$~s are routinely achieved on targets with similar characteristics \citep{nascimbeni2011}.

As for the occultation, the second epoch has to rely on space-based instruments in the NIR/MIR spectral region, the only capable of timing the $\sim 1900$~ppm eclipse with the required precision. After the retirement of Spitzer, the first viable option is with JWST \citep{beichman2014}: a single occultation observed with the NIRSpec instrument will be able to constrain its timing to better than a few seconds, at epoch >2022.5 (that is, after the foreseen start of JWST cycle 1). Indeed, the full-phase observation of WASP-14 b with NIRSpec was once considered as a  Science Operations Design Reference Mission (SODRM) program, but eventually not included in any ERS (Early Release Science; \citealt{bean2018}) or GTO program. It is likely, however, that it will be included as an ordinary program during the first cycles, either to study its atmosphere or specifically to detect the GR effect through a dedicated proposal.  A second option is with ARIEL \citep{tinetti2018}, with an expected launch date in 2029; WASP-14 b is already included in the Mission Reference Sample \citep{zingales2018} in Tier 3 (the most intensively monitored sample), specifically to gather repeated observations of its occultations. Also in this case, the expected timing error will be better than a few seconds, and at an even more distant epoch ($>20$~yr) after the first epoch set by \citealt{blecic2013} with Spitzer.

\begin{table*}
\centering
 \begin{tabular}{c c c c c c c} 
 \hline
 Name & T (y) & $|\Delta \tau_{GR}|$ (s) & ED 3.6 $\mu$m (ppm) & ED 4.5 $\mu$m (ppm) & V (mag) & References \\ [0.5ex] 
 \hline\hline
 WASP-14 b & 12 & 130 & 1816 $\pm$ 67 & 2161 $\pm$ 88 & 9.75 & \cite{joshi2009} \\ 
 HAT-P-33 b & 14 & 220 & 1603 $\pm$ 127 & 1835 $\pm$ 199 & 11.89 & \cite{hartman2011} \\ 
 HAT-P-30 b & 11 & 40 & 988 $\pm$ 168 & 1057 $\pm$ 145 & 10.42 & \cite{johnson2011} \\ 
 KELT-2A b & 9 & 50 & 650 $\pm$ 38 & 678 $\pm$ 47 & 8.77 & \cite{beatty2012} \\
 XO-3 b b & 8 & 50 & --- & 1580 $\pm$ 36 & 9.8 & \cite{wong2014} \\
 [1ex] 
 \hline
 \end{tabular}
\caption{The best candidates for detecting the GR precession rate in eccentric HJ with a well defined secondary eclipse. The second column (T) contains the length of the observational time-span for each target; the third column lists an approximate computation (ignoring uncertainties) of the expected total $\Delta \tau$, based upon the best values of their masses and orbital parameters, to give a grasp of the magnitude of the effect; the fourth and the fifth columns give the depths of the secondary eclipses in two wavelengths (3.6 $\mu$m and 4.5 $\mu$m) from \citet{garhart2020}; the depth 4.5 $\mu$m of the secondary eclipse of XO-3 b is from \citet{wong2014}; the sixth columns contains the visual magnitudes of the parent stars.}
\label{table:2}
\end{table*}

\section{Conclusions and discussion}

In this work we have presented a convenient way to detect the GR precession rate of transiting HJs with a secondary eclipse sufficiently deep to measure with high precision ($\lesssim$ 1 min) the time interval between the mid-time of the primary transit and the mid-time of the occultation. Our method is based on measuring the above mentioned time interval, $\Delta \tau$, which is a differential measurement and thus it is not affected by the accumulation of uncertainty in determining the epochs of these events. We also accounted for the tidal and rotational effects, as they are, among the classical sources of precession, the ones with the greatest rates, comparable with the GR precession rate.

In order to test our method, we propose the HJ WASP-14 b as the most promising target for future observations. We demonstrated that it is possible to detect GR precession rates already with current space (TESS) and ground-based facilities and also with future space-based missions (JWST and ARIEL). We derived an expected $\Delta \tau$ $\sim$ 4 min, including both GR and classical effects, for WASP-14 b, considering an observational time-span of 12 yr. This time-span corresponds to the time elapsed since the first photometric observations of WASP-14 and the present year. However, we provide also a way of computing the expected $\Delta \tau$ as a function of an arbitrary observational time-span. Since GR precession is a cumulative effect, a longer observational time-span will lead to more accurate measurements of $\Delta \tau$.

In this perspective, future IR observations and especially high precision space-based observations (JWST) of WASP-14 b and of the other targets listed in Tab.~\ref{table:2} will be of great importance, because they will make it possible to test the predictions of GR in the field of exoplanets with a growing level of accuracy.

\section*{Acknowledgements}

LBo and GLa acknowledge the funding support from Italian Space Agency (ASI) regulated by ``Accordo ASI-INAF n. 2013-016-R.0 del 9 luglio 2013 e integrazione del 9 luglio 2015 CHEOPS Fasi A/B/C''. GLa acknowledges support by CARIPARO Foundation, according to the agreement CARIPARO-Universit{\`a} degli Studi di Padova (Pratica n. 2018/0098).

\section*{Data Availability}

No new data were generated or analysed in support of this research.




\bibliographystyle{mnras}
\bibliography{references}







\bsp	
\label{lastpage}
\end{document}